\documentclass[a4paper]{jpconf}
\usepackage{graphicx}
\begin{document}

\title{Neutron Star Crust in Strong Magnetic Fields}

\author{Rana Nandi and Debades Bandyopadhyay}

\address{Astroparticle Physics and Cosmology Division,
Saha Institute of Nuclear Physics, 1/AF Bidhannagar, 
Kolkata-700064, India}

\ead{debades.bandyopadhyay@saha.ac.in}

\begin{abstract}
We discuss the effects of strong magnetic fields through Landau quantization of
electrons on the structure and
stability of nuclei in neutron star crust. In strong magnetic fields, this  
leads to the enhancement of the electron number density with respect to the 
zero field case. We obtain the sequence of equilibrium nuclei of the outer 
crust in the presence of strong magnetic fields adopting most recent versions 
of the experimental and theoretical nuclear mass tables. For $B\sim 10^{16}$G,
it is found that some new nuclei appear in the sequence and some nuclei 
disappear from the sequence compared with the zero field case. 

Further we investigate the stability of nuclei in the inner crust in the 
presence of strong magnetic fields using the Thomas-Fermi model. The 
coexistence of two phases of nuclear matter - liquid and gas, is considered in 
this case. The proton number density is significantly enhanced in 
strong magnetic fields $B\sim 10^{17}$G through the charge neutrality. 
We find nuclei with larger mass number in the presence of 
strong magnetic fields than those of the zero field. These results might have 
important implications for the transport properties of the crust in magnetars.   

\end{abstract}

\section{Introduction}
The discovery of a new class of neutron stars with very strong magnetic fields 
called magnetars has greatly enhanced the interest in the study of neutron star
properties in the presence of strong magnetic fields \cite{lai}. 
Their surface magnetic fields could be $\geq 10^{15}$, as predicted by 
observations on soft gamma-ray repeaters and anomalous x-ray pulsars 
\cite{vas,kouvel98}. Such strong magnetic fields might be generated by dynamo
processes in newly born neutron star \cite{thomp93}. In the core region the 
magnetic field may be even higher, the limiting value ($B_{max}$) is obtained 
by the scalar virial theorem \cite{lai91}. 
For a typical neutron star ($M=1.5 M_{\bigodot},R=15$ km) this value is 
$B_{max}\sim 10^{18}$ G. Such high magnetic fields can have significant effects
on the equilibrium nuclear composition and equation of state in neutron star
crust and interior \cite{lai91,ban}.

Nonmagnetic equilibrium composition
and equation of state for the outer crust was reported in a seminal paper by 
Baym, Pethick and Sutherland (BPS) \cite{bps}. Outer crust 
contains nuclei arranged in a body-centered (bcc) lattice immersed in a gas of 
free electrons which are relativistic above the density $\rho\sim 10^7 
$g cm$^{-3}$. Though the lattice effect is small on the equation of state, it 
changes the equilibrium nucleus to a heavier one and lowers the total energy of 
the system by reducing the coulomb energy of the nucleus. At $\rho\sim 10^4$ 
g cm$^{-3}$, $^{56}$Fe is present as the equilibrium nucleus, but with 
increasing density equilibrium nuclei become more and more neutron rich through 
electron capture process. At a density $\rho \simeq 4\times10^{11}$
g cm$^{-3}$ neutrons begin to drip out of nuclei - this is called neutron drip 
point. The inner crust begins from here. In the inner crust nuclei are immersed
in a neutron gas as well as a uniform background of electrons. Nuclei are 
also arranged in a lattice in the inner crust. The composition and the equation
of state of the inner crust were earlier calculated by different groups 
\cite{bbp,neg}. 

The composition and equation of state of the outer crust of 
nonaccreting cold neutron stars in the presence of strong magnetic fields were
first studied by Lai and Shapiro \cite{lai91}. 
In the presence of a magnetic field the motion of electrons perpendicular to 
the field get quantized into Landau orbitals. This causes the 
electron density to change which in turn modifies the coulomb energy. If the 
magnetic field is very strong then electrons occupy only the low-lying Landau 
levels and it may affect the sequence of nuclei and the equation of state as 
well as any nonequilibrium $\beta$-processes \cite{lai91}. However, there is
no calculation of the inner crust composition and equation of state in the
presence of magnetic fields.

This paper is organised in the following way. We revisit the magnetic BPS 
\cite{lai91} adopting recent experimental and theoretical nuclear mass tables 
in Section II. The inner crust in strong magnetic fields is discussed in 
Section III. We conclude in Section IV.

\section{Magnetic BPS Model} 

We revisit the BPS model to find the sequence of equilibrium nuclei and 
calculate the equation state of the outer crust in the presence of strong 
magnetic fields $B\sim 10^{16}$G \cite{lai91}. In this calculation, we include
the finite size effect in the lattice energy and adopt recent experimental and
theoretical mass tables. Nuclei are arranged in a bcc lattice in the 
outer crust. The Wigner-Seitz (WS) approximation is adopted in this 
calculation. Each lattice volume is replaced by a spherical cell and  contains 
one nucleus at the center. Each cell is taken to be charge neutral such that
Z  number of electrons presents in it, where Z is the nuclear charge. The 
Coulomb interaction between cells is neglected. To find an equilibrium nucleus
(A,Z) at a given pressure P one has to minimize the Gibbs free energy per 
nucleon with respect to A and Z. The total energy density is given by
\begin{equation}
 E_{tot} = n_N (W_N+ W_L)+ \varepsilon_e(n_e)~.
\end{equation}
The energy of the nucleus (including rest mass energy of nucleons) is
\begin{equation}
W_N = m_n (A - Z) + m_p Z - bA~,
\end{equation}
where $b$ is the binding energy per nucleon.
Experimental nuclear masses are obtained from the atomic mass table 
compiled by Audi, Wapstra and Thibault (2003) \cite{audi03}. For the rest of 
nuclei we use the theoretical extrapolation of M\"{o}ller et al (1995) 
\cite{moller95}. $W_L$ is the lattice energy of the cell and is given by
\begin{equation}
  W_L= -\frac{9}{10}\frac{Z^2e^2}{r_C}\left(1-\frac{5}{9} 
\left(\frac{r_N}{r_C}\right)^2\right)~.
\end{equation}
Here $r_C$ is the cell radius and $r_N\simeq r_0A^{1/3}$ ($r_0\simeq$1.16 fm) 
is the nuclear radius. The first term in $W_L$ is the lattice energy for point 
nuclei and the second term is the correction due to the finite size of the 
nucleus (assuming a uniform proton charge distribution in the nucleus). Further
$\varepsilon_e$ is the electron energy density and P is the total pressure 
given by
\begin{equation}
 P=P_e+\frac{1}{3}W_Ln_N.
\end{equation}
The nucleon number density $n_N$ is related to the baryon number density $n_b$ 
as
\begin{equation}
 n_b=A n_N,
\end{equation}
and the charge neutrality condition gives
\begin{equation}
 n_e=Zn_N.
\end{equation}

\begin{figure}[h]
\begin{center}
\includegraphics[width=7cm,angle=-90]{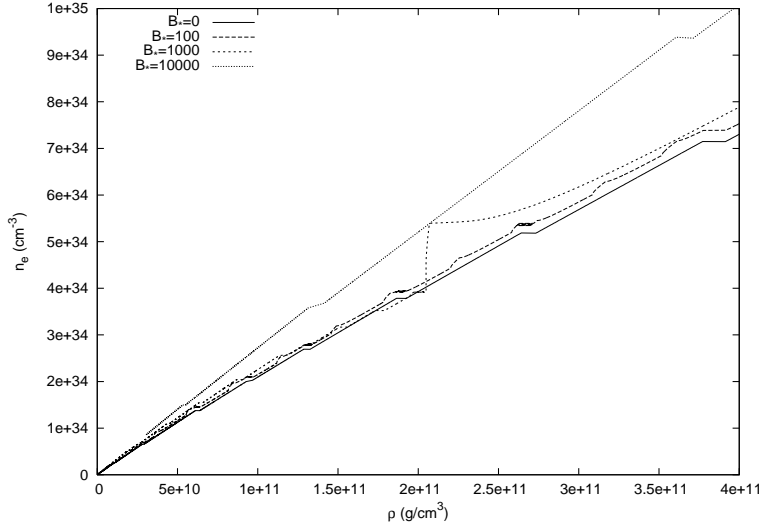}\hspace{2pc}%
\end{center}
\caption{\label{label}Electron number density is plotted with mass density for 
different magnetic field strengths.}
\end{figure}

In the presence of a magnetic field, the electron motion is Landau quantized in
the plane perpendicular to the field. We take the magnetic field 
($\overrightarrow{B}$) along Z-direction and assume that it is uniform 
throughout the outer crust. If the field strength exceeds a critical value 
$B_c=m_e^2/e\simeq 4.414\times 10^{13}$G, then electrons become relativistic. 
The energy eigenvalue of relativistic electrons in a quantizing magnetic field 
is given by
\begin{equation}
 E_e(\nu,p_z)=\left[p_z^2+m_e^2(1+2\nu B_{*})\right]^{1/2}~,
\end{equation}
where $p_z$ is the Z-component of momentum, $B_{*}=B/B_c$, and $\nu$ is 
the Landau quantum number.

The number density of electrons in a magnetic field is calculated as
\begin{equation}
 n_e=\frac{eB}{2\pi^2}\sum_0^{\nu_{max}}g_{\nu}p_{f_{e}}(\nu)~,
\end{equation}
where the spin degeneracy $g_\nu=1$ for $\nu=0$ and $g_\nu = 2$ for all other
Landau levels. 
The maximum Landau quantum number $\nu_{max}$ is given by
\begin{equation}
\nu_{max}=\frac{\mu_e^2-m_e^2}{2eB}~,
\end{equation}
where $\mu_e$ is the electron chemical potential. 
Energy density of electrons is given by,
\begin{eqnarray}
\varepsilon_e=\frac{eB}{2\pi^2}\sum_0^{\nu_{max}}g_\nu \int_0^{p_{f_{e}}(\nu)}
E_e(\nu,p_z)dp_z~,
\end{eqnarray}
and the pressure of the electron gas is 
\begin{eqnarray}
 P_e&=&\mu_e n_e-\varepsilon_e~. 
\end{eqnarray}

At a fixed pressure $P$, we minimize the Gibbs free energy per nucleon 
\begin{eqnarray}
g = \frac{E_{tot} + P}{n_b} = \frac{W_N + 4/3 W_L + Z \mu_e}{A}~,
\end{eqnarray}
varying $A$ and $Z$ of a nucleus.

Now we present our results for several values of magnetic fields 
$B_{*} = B/B_c$ = 0, 100, 1000 and 10000 where $B_c = 4.414 \times 10^{13}$G.
In Fig. 1, electron number density is plotted with mass density for
the above mentioned values of magnetic fields. For $B_{*} <$ 1000, large numbers
of Landau levels are populated. Consequently there is no significant change in 
the electron number density compared with that of the field free case. But 
there is a jump in the electron number density for $B_{*} = 10^3$ case when 
only zeroth Landau level is populated at $\sim 2 \times 10^{11}$ g/cm$^{3}$. On
the other hand, for $B_{*} = 10^4$, only zeroth Landau level is populated over 
the whole mass density range. In this case, the phase space modification due to
the Landau quantization leads to the strong enhancement of electron number 
density 
with respect to the zero field case.

\begin{figure}[h]
\begin{center}
\includegraphics[width=7cm,angle=-90]{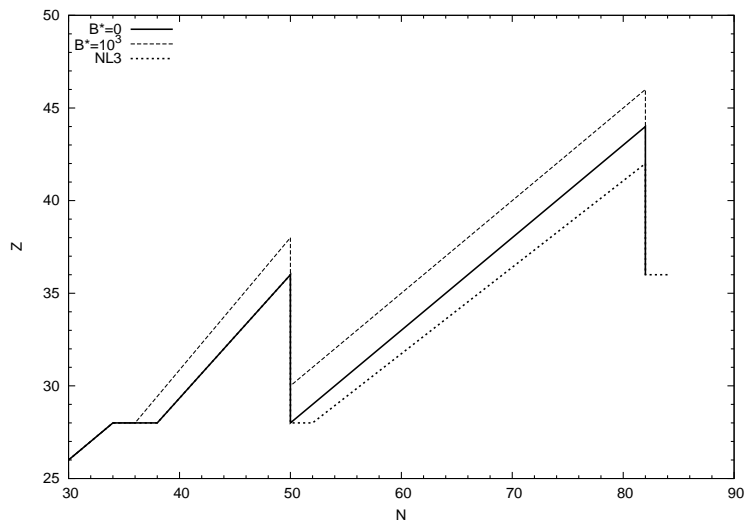}\hspace{2pc}%
\end{center}
\caption{\label{label}Proton number is shown as a function of neutron number 
for different theoretical nuclear models with and without magnetic fields.}
\end{figure}

We obtain the sequence of equilibrium nuclei minimizing the Gibbs free energy 
per nucleon for $B=0$ and $B=10^{16}$G. These results are obtained with and 
without the lattice energy correction. We compare our findings at
$B_{*} = 10^3$ with those of zero field case and find that some new 
equilibrium nuclei such as $^{88}_{38}$Sr and $^{128}_{46}$Pd appear and some 
nuclei such
as $^{66}$Ni and $^{78}$Ni disappear in the presence of the magnetic field. 
Further we note that the maximum density ($\rho_{max}$) upto which an 
equilibrium nucleus can exist, increases as the field strength
increases. As a result the neutron drip point is shifted from 
$4.34\times10^{11}$ g/cm$^3$ in zero field to $4.92\times10^{11}$ g/cm$^3$ in
the strong magnetic field. 
The lattice energy correction influences our results in strong 
magnetic fields. It is worth mentioning here that our results are different 
from those of earlier calculation \cite{lai91} because we have adopted most 
recent experimental and theoretical nuclear mass tables. For $B=0$, we note 
that $^{122}_{38}$Sr which was present in the previous calculation \cite{lai91}
is absent in our case. A comparison of our results at zero field with those of
Lai and Shapiro \cite{lai91} shows that equilibrium nuclei have higher 
$\rho_{max}$ in the former case. Further we performed our calculation at 
higher magnetic fields than the previous calculation \cite{lai91}.

Figure 2 displays the proton number as a function of neutron number. We compare
sequences of equilibrium nuclei calculated with experimental nuclear masses 
from atomic mass table of Audi, Wapstra and Thibault \cite{audi03} and 
theoretical nuclear mass models for those nuclei which are not listed in the 
experimental mass table. We use the theoretical model of M\"oller et al.
\cite{moller95} and relativistic mean field model with NL3 set 
\cite{hempel,ben}. 
It is evident from the figure that for zero magnetic field our calculation of 
equilibrium nuclei initially agrees with those of the relativistic model 
calculation because nuclear masses are obtained from the
experimental mass table. However, both calculations for zero magnetic field 
differ considerably beyond N=50 due to differences in theoretical mass tables 
used here. 

\begin{figure}[h]
\begin{center}
\includegraphics[width=7cm,angle=-90]{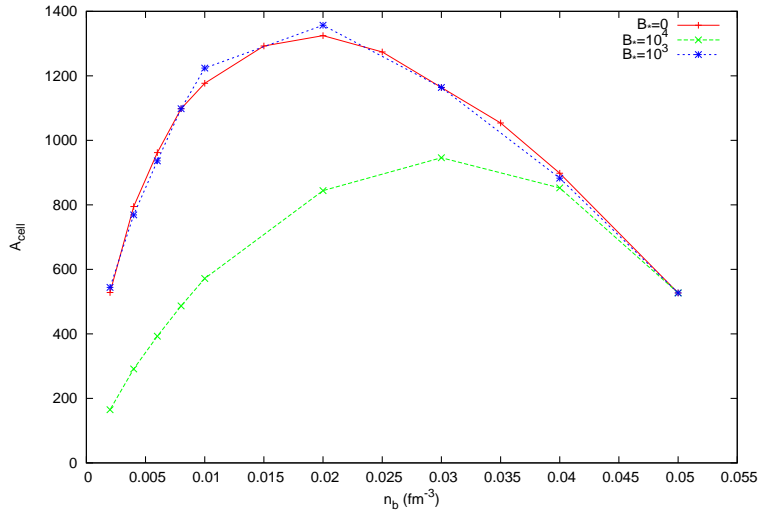}\hspace{2pc}%
\end{center}
\caption{Mass number is displayed as a function of average 
baryon density.}
\end{figure}

\section{Inner Crust in Magnetic Fields} 
   
Next we investigate the stability of nuclei in the inner crust in the presence
of strong magnetic fields using the Thomas-Fermi model. In this case nuclei are
immersed in a neutron gas as well as in a uniform background of electrons.
The coexistence of liquid and gas phases of nuclear matter is considered here. 
In the inner crust, nuclei are also arranged in a lattice. The Wigner-Seitz 
approximation is adopted in this calculation. However, the 
spherical cell in which protons and neutrons coexist does not define a 
nucleus. The nucleus is realised after subtraction of the gas part from the 
cell. This is done following the procedure of Bonche, Levit and Vautherin 
\cite{bon}. In
this case, the density profile of liquid plus gas system and that of the gas
are obtained in a self-consistent way and the liquid part which ultimately
becomes the nucleus, is calculated as the difference of two solutions. We 
assume $\beta$-equilibrium $\mu_n = \mu_p + \mu_e$. 
Electrons are affected by quantizing magnetic fields like the outer crust. 
The proton density in the
cell is affected by magnetic fields through the charge neutrality condition.
We minimise the free energy which is a function of average baryon density 
($n_b$) and proton fraction ($Y_p$), in the cell
\begin{equation}
{\cal{F}}(n_b,Y_p) = \int [{\cal{H}} + \varepsilon_c + \varepsilon_e] d{\bf r}~,
\end{equation}  
where $\cal{H}$ is nuclear energy density functional, $\varepsilon_c$ is the 
coulomb 
energy density and $\varepsilon_e$ is the energy density of electrons. The 
nuclear
energy density is calculated using SKM$^*$ nucleon-nucleon interaction
\cite{sil}. We perform this calculation for temperature T=0. 

Now we present our results related to the inner crust. We fix the
average baryon density $n_b$ and achieve the $\beta$-equilibrium changing 
the proton fraction $Y_p$. Next we find the minimum
of the free energy varying the cell size. We perform this calculation 
for $B=0$ and $B_{*} = 10^3, 10^4$. We do not find any appreciable difference
in our results for $B=0$ and $B_* = 10^3$. However, the cell size is 
appreciably reduced in case of $B_* = 10^4$ . In this case only zeroth Landau 
level is
populated below the density 0.06 fm$^{-3}$. This leads to the large enhancement
of
proton fraction in the cell through the Landau quantization of electrons. Above
this density, electrons populate finite number of Landau levels and the cell 
size and proton fraction in the magnetic field case become equal to those of 
the zero field. The size of the cell as well as proton fraction decrease as 
density increases. 

Figure 3 displays mass number (A$_{cell}$) of nuclear cluster as a function of 
average 
baryon density for $B=0$ and $B_{*} = 10^3, 10^4$. These are obtained after 
minimising the free energy at each density point. It is noted that at any 
density point the mass number of the nuclear cluster in magnetic field 
$B_* = 10^4$ is lower than the corresponding nuclear cluster at zero magnetic 
field. Further we obtain the liquid part of the cell  after 
subtraction of the gas part and define this as a nucleus. We find nuclei 
with larger mass number in the presence of a strong magnetic field 
$\sim 10^{17}$G compared with the zero field case. Further nuclei are more 
bound in strong magnetic fields than the corresponding nuclei in $B=0$ at the 
same baryon density.

\section{Summary and Conclusions}
We have revisited the BPS model of outer crust in the presence of strong 
magnetic
fields $\sim 10^{16}$G or more using the recent experimental mass table. 
Further we have included the correction in the lattice energy due to the finite
size of a nucleus. For zero magnetic field case, it is noted that maximum 
densities for heavier nuclei in this calculation are higher than those of the 
previous calculation \cite{lai91}. In the presence of a strong magnetic field, 
there are modifications in the sequence of nuclei compared with the zero field 
case. We have further investigated the inner crust in the presence of strong
magnetic fields. For $B=10^{17}$G, it has been observed that proton fraction 
is enhanced at lower densities and mass numbers of nuclear clusters 
after subtracting the gas part are higher than those of the zero field.
It would be worth investigating the implications of our results for the 
transport properties such as thermal and electrical conductivities and
shear viscosity of the crust in magnetars.

\section{Acknowledgment}
We acknowledge the support of the Alexander von Humboldt Foundation 
under the Research Group Linkage Programme.

\end{document}